\documentclass[runningheads]{llncs}
\usepackage{graphicx}

\usepackage{tikz}
\usepackage{comment}
\usepackage{amsmath,amssymb} 
\usepackage{color}
\usepackage{booktabs}
\usepackage{algorithm,algpseudocode}
\usepackage{tablefootnote}
\usepackage{amsmath}
\usepackage{xcolor}

\usepackage[accsupp]{axessibility}  

\begin{document}
\pagestyle{headings}
\mainmatter
\def\ECCVSubNumber{001}  

\title{Synthetic Dataset Generation for Privacy-Preserving Machine Learning} 

\titlerunning{Synthetic Dataset Generation}
%
\author{Efstathia Soufleri, Gobinda Saha, Kaushik Roy}
%
\institute{ECE, Purdue University, USA}
\maketitle

\begin{abstract}
Machine Learning (ML) has achieved enormous success in solving a variety of problems in computer vision, speech recognition, object detection, to name a few. The principal reason for this success is the availability of huge datasets for training deep neural networks (DNNs). However, datasets can not be publicly released if they contain sensitive information such as medical or financial records. In such cases, data privacy becomes a major concern. Encryption methods offer a possible solution to this issue, however their deployment on ML applications is non-trivial, as they seriously impact the classification accuracy and result in substantial computational overhead.
Alternatively, obfuscation techniques can be used, but maintaining a good balance between visual privacy and accuracy is challenging. In this work, we propose a method to generate secure synthetic datasets from the original private datasets. In our method, given a network with Batch Normalization (BN) layers pre-trained on the original dataset, we first record the layer-wise BN statistics. Next, using the BN statistics and the pre-trained model, we generate the synthetic dataset by optimizing random noises such that the synthetic data match the layer-wise statistical distribution of the original model.       
We evaluate our method on image classification dataset (CIFAR10) and show that our synthetic data can be used for training networks from scratch, producing reasonable classification performance.\footnote{Work in Progress}

\end{abstract}

\section{Introduction}
Machine Learning (ML) has been integrated with great success in a wide range of applications such as computer vision, autonomous driving, speech recognition, natural language processing, object detection and so on. The availability of large datasets and advancements in techniques for training deep neural network models have played integral roles towards such success. Moreover, the cloud providers offer various Machine Learning as a Service (MLaaS) platforms such as Microsoft Azure ML Studio \cite{microsoft_citation}, Google Cloud ML Engine \cite{google_citation}, and Amazon Sagemaker \cite{amazon_citation} etc., where computational resources is provided for running ML workloads. For such cloud-based computing, the ML algorithms are either provided by the users or selected from the standard ML algorithm libraries~\cite{zhang2018privacy}, whereas the datasets are usually shared to cloud by the users to meet application-specific requirements. However, it might not be always possible to share private data to the cloud if they contain sensitive information such as medical or financial records or the user may need to follow certain terms and regulations that forbid public release of the data. 

Thus, it is essential to protect the privacy of the training data before publicly releasing them, as this would offer numerous advantages. First, it will be beneficial for the research community - scientists could open source their data without privacy concerns and consequences. Access to the data will facilitate researchers to reproduce experiments from other studies and hence, transparency will be promoted. Additionally, by combining data from different sources can lead to better models that can be built. Moreover, data can potentially be traded in data markets, where protection of sensitive information is of utmost importance \cite{triastcyn2018generating}. Overall, collaboration between users will be facilitated and this will help in a broader way towards advancements in ML research. 

In literature, several proposals have been suggested to ensure visual privacy of the data. Among those methods, a popular approach is data encryption \cite{bost2014machine,sirichotedumrong2019privacy,aono2016scalable,bonte2018privacy,crawford2018doing,kim2018secure}, where the user encrypts the data before sending to the cloud. This is considered to be a highly successful method and has demonstrated exceptional results for protecting the data privacy \cite{graepel2012ml,nandakumar2019towards}. 
However, this method is computationally expensive making it prohibitive for wide adaptation in ML applications \cite{jayaraman2019evaluating}. Alternatively, researchers have suggested several image-obfuscation techniques for visual privacy preservation~\cite{triastcyn2018generating}. These techniques include image blurring \cite{neustaedter2006blur}, mixing \cite{inoue2018data,zhang2017mixup}, adding noise \cite{zhang2018privacy}, pixelizing \cite{raynal2020image} etc., which allow the model to be trained with the obfuscated images and achieve good accuracy-privacy trade-off. However, when the images are highly obfuscated, though visual privacy is well ensured, the performance of the network might drop beyond the desirable point. 

In this work, we introduce an algorithm for generating secure synthetic data from the original private data. Specifically, as inputs, our method requires the original data and a network with Batch Normalization (BN) layers pre-trained on these data. From this network we record the BN statistics - the running mean and running variance - from each layer. Then, we initialize the synthetic data as Gaussian noise and optimize them to match the recorded BN statistics \cite{cai2020zeroq}. 
Upon generation of the synthetic dataset, they can be publicly released and used for training DNNs from scratch.
We evaluate our methodology on CIFAR10 \cite{krizhevsky2009learning} image classification datasets. We train a network (ResNet20) from scratch on synthetic CIFAR10 images and obtain up to 61.79\% classification accuracy. We show that such classification performance depends on number of optimization steps used for synthetic data generation. For longer optimization steps classification performance increases, however the generated images look more real-like sacrificing some data privacy.   

\section{Related Work}
One approach to tackle the problem of maintaining data privacy is to apply noise on the input or the gradients of the NN in a differentially private manner \cite{huang2019dp,papernot2018scalable,papernot2016semi,fredrikson2014privacy}.
Differentially Private Stochastic Gradient Descent (DPSGD) \cite{abadi2016deep} method clips and adds noise in the gradients of the model during training in order to prevent the model to leak information regarding the data. Authors in \cite{fan2019differential,lee2019synthesizing} apply noise directly to the raw input data. The above approaches provide strong privacy guarantees at the detriment of accuracy.    

Image obfuscation is a technique that visually distorts parts or the entire image \cite{zhang2018privacy,fan2019practical,raynal2020image,triastcyn2018generating,li2021deepblur,poller2012robust}. Thus, they are able to hide the sensitive information and securely release the images for ML applications. Zhang et al. \cite{zhang2018privacy} add noise in the image, train an NN and then evaluate it against model inversion attack \cite{fredrikson2015model}, model classification attacks \cite{ateniese2015hacking}, membership inference attack \cite{shokri2017membership} and model memorization attack \cite{song2017machine}. Authors in \cite{raynal2020image} apply blurring, pixelization and other forms of distortion on the image. They perform a study that evaluates which metrics are aligned with human perception as well as how much distortion is required such that the image can not be classified correctly. They evaluate various image quality metrics and conclude that Structural Similarity Index Measure (SSIM) and Haar wavelet-based perceptual similarity index (HaarPSI)
align better with the human perception \cite{raynal2020image}. Researchers in \cite{triastcyn2018generating} present a method for generating Differential Private (DP) synthetic data. This method adds noise in the data generated by a Generative Adversarial Network (GAN). However, the images might reveal information about the original images or might not maintain the utility of the data resulting in high accuracy degradation.  

Another direction is the privacy-preserving data release \cite{triastcyn2018generating,lee2019synthesizing,alkhelaiwi2021efficient,liu2021machine}. This approach offers numerous advantages such as any model can be trained flexibly on the released data, combined data from different sources can be used to build stronger models etc. Our proposal falls in this category. We suggest a method for generating the synthetic images that preserves the statistical properties of the original images. Note, that our proposed method can potentially be used in conjunction with the aforementioned DPSGD method, adding one more layer of protection against attacks. 

\begin{figure}[t]
\centering
\includegraphics[width=0.9\textwidth]{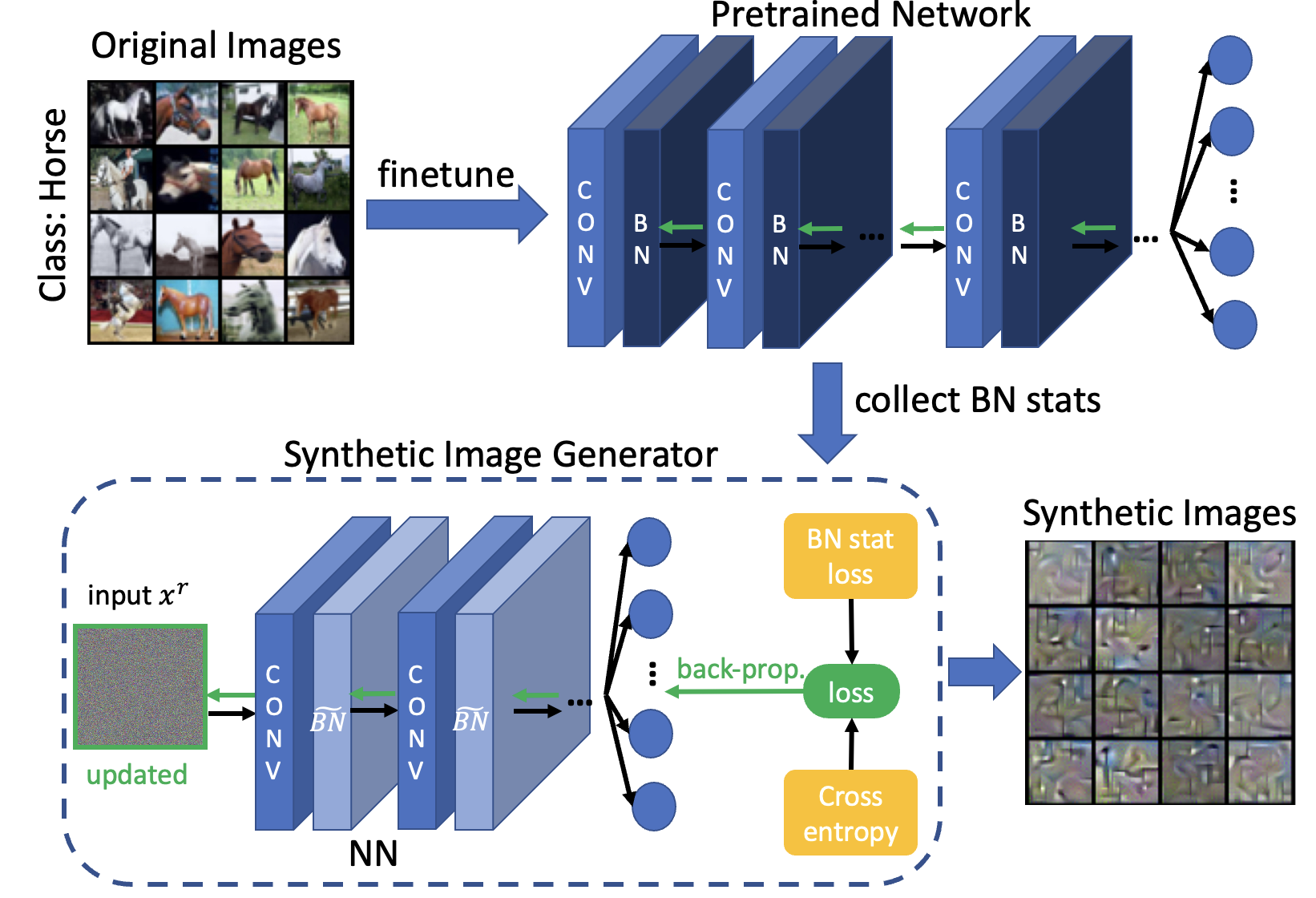} 
\caption{Overview of the proposed method. From a pre-trained network the BN statistics (running mean \& variance) are collected. The Synthetic Image Generator optimizes random noise such that the synthetic images have a statistical distribution that closely matches the BN statistics of the pre-trained NN. The BN stat loss is: $\sum_{i=1}^{L} ||\tilde{{\mu_i}^r} - \mu_i||_2^{2} + ||\tilde{{\sigma_i}^r} - \sigma_i||_2^{2}$ between the collected BN statistics ($\mu_i$, $\sigma_i$) and the mean and variance ($\tilde{{\mu_i}^r}$, $\tilde{{\sigma_i}^r}$) of the synthetic images ($x^r$) activations and the cross entropy loss is: $f(labels, NN(x^r))$ between the actual label and output of the NN on synthetic data.}
\label{fig:methodology overview}
\vspace{-25pt}
\end{figure}

\section{ Synthetic Image Generation Method}


In this section, we describe the methodology for generating the synthetic data. Figure \ref{fig:methodology overview} gives an overview of the proposal. A neural network (NN) with BN layers is first trained (or finetuned) on the original images. 
Then from the pre-trained model we collect the layerwise BN statistics, namely the running mean ($\mu$) and the variance ($\sigma$). Initially, we define a batch of the synthetic images as random Gaussian noise. These images are optimized based on a loss function such that it resembles the BN statistics of the corresponding model pre-trained on original dataset. To generate synthetic images ($x^r$), we optimize the following loss function:      
\begin{align}
\label{eqn:loss function}
  \min_{x^r}  \sum_{i=1}^{L} ||\tilde{{\mu_i}^r} - \mu_i||_2^{2} + ||\tilde{{\sigma_i}^r} - \sigma_i||_2^{2} + f(labels, NN(x^r))
\end{align}
where $\mu_i$ and $\sigma_i$ are the running mean and variance at layer $i$ of the NN recorded from the pre-trained model, whereas $\tilde{{\mu_i}^r}$ and $\tilde{{\sigma_i}^r}$ are the mean and variance of activations of the synthetic images ($x^r$) at layer $i$. $L$ is the number of layers of the pre-trained network, $NN(x^r)$ is the output of the NN when $x^r$ is fed as an input, and $f$ denotes the cross-entropy loss between the actual labels and the output of the network. Essentially, this optimization process aims to generate synthetic images ($x^r$) such that, when fed in the NN, produces layer-wise statistical distributions that closely matches distributions of the pre-trained model. Each batch of synthetic images is optimized for $k$ steps, where $k$ is a hyperparameter. The synthetic images obtained after such optimization steps can be released publicly to enable privacy-preserving ML applications. The steps for generating these images in the form of pseudo-code is given in Algorithm~\ref{syn_algo}.

\begin{algorithm*}[!t]
  \caption{: Synthetic Image Generation}\label{syn_algo}
  \begin{algorithmic}[1]
  \footnotesize
    \State \textbf{Input}: Original Images with $C$ classes; 
        NN with $L$ BN layers pre-trained on the original images; $k$ is the number of optimization steps for the synthetic image generation; $b_s$ is the batch size for synthetic data; $N$ is the number of total synthetic images; iter is $\frac{N}{b_s}$
    \State \textbf{Output}: Synthetic Images 
    \State \textbf{Generation of Synthetic Images:}
    \For{i = 1,2, ..., iter} 
        \State Record the BN statistics ($\mu, \sigma$)
        \State Generate random data from Gaussian $x^r$ (batch of images of size $b_s$)
        \State Assign labels to $x^r$ 
        \For {j = 1, 2, ..., k} 
            \State Forward Propagate the NN($x^r$)
            \State Record the BN statistics ($\tilde{{\mu_i}^r}$, $\tilde{{\sigma_i}^r}$) of NN($x^r$)
            \State Compute the loss from eq. \ref{eqn:loss function}
            \State Backward Propagate and update $x^r$
        \EndFor
        \State Store the $x^r$ 
    \EndFor
     \State \textbf{Return} synthetic images of all the classes
  \end{algorithmic}
\end{algorithm*}

\section{Experiments and Results}

\subsection{Experimental Setups}
We evaluate our proposed method on image classification datasets - CIFAR10 \cite{krizhevsky2009learning}. We use ResNet20 \cite{he2016deep} model. We implemented our experiments using the PyTorch framework \cite{NEURIPS2019_9015} on a system with a NVIDIA GTX 1080 Ti GPU. For synthetic data generation, we use Adam optimizer \cite{kingma2014adam} with learning rate = 0.1 , $\beta = (0.9, 0.999)$, and batch size, $b_s$ = 250 for $k=\{250,500,1000\}$ iterations per batch. In total we generate $N=50000$ synthetic images. For training the ResNet20 model on the synthetic dataset, we use SGD optimizer \cite{kiefer1952stochastic} with learning rate = 0.1 and learning rate scheduler with Cosine Annealing. The model is trained for 200 epochs with batch size = 128 for CIFAR10. The pre-trained ResNet20 on CIFAR10 was obtained from the pytorchcv open source library \cite{pypi}.

\subsection{Results and Analyses}
\begin{figure*}[!ht]
\begin{centering}
  \includegraphics[width=0.80\textwidth,keepaspectratio,page=3]{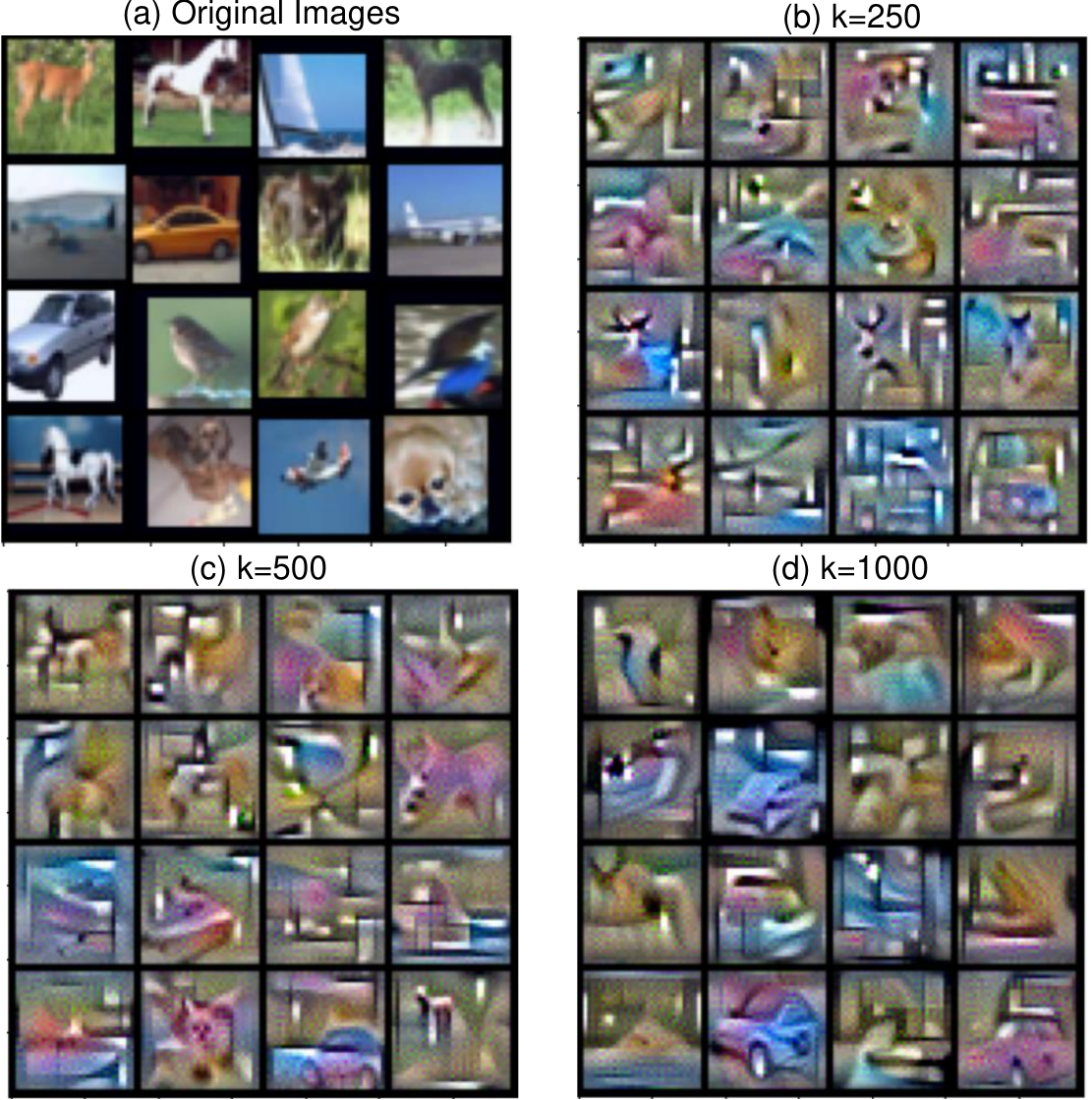}
  \caption{(a) Original CIFAR10 data samples used to pre-train the model for synthetic data generation. Synthesized data samples from the pre-trained ResNet20 model using (b) $k=250$ (c) $k=500$ and (d) $k=1000$ data optimization steps per batch.} 
\label{fig:fig2}
\end{centering}
\vspace{-15pt}
\end{figure*}

In Figure~\ref{fig:fig2}(b)-(d) we show the synthetic data samples generated with our approach. Original CIFAR10 images used to pre-train the ResNet20 model is also provided in Figure~\ref{fig:fig2}(a) for visual comparison reference. From these figures we can see that, as more optimization steps ($k$ values) are used for data generation, the generated synthetic data reveals more visual features of the underlying classes. In Table~\ref{table:accuracy_cifar10}, we provide classification accuracy of the ResNet20 model trained from scratch on synthetic images. Here, we find that data generated with higher $k$ values has the highest accuracy. For instance we obtained accuracy of 61.79\% from synthetic data generated with $k=1000$. Thus, accuracy-privacy trade-off can be obtained in our method by tuning the values of $k$, where with larger $k$ we can obtain better accuracy sacrificing some visual privacy. In Table~\ref{table:accuracy_cifar10} we also provide accuracy of a ResNet20 model trained on original CIFAR10 dataset. Here we obtained $\sim30\%$ better accuracy than the synthetic image based training. These results show the necessity for better data optimization and design space explorations before using synthetic dataset as drop-in replacement for original images in high performance applications. 

\setlength{\tabcolsep}{4pt}
\begin{table}[h]
\begin{center}
\caption{Classification results (Top-1 accuracy) for ResNet20 networks trained on synthetic and original CIFAR10 images. Here results for synthetic images are shown for varying length of data optimization steps, $k$ values. The synthetic images were generated using the BN statistics of the pre-trained ResNet20 networks on CIFAR10.}
\vspace{-10pt}
\label{table:accuracy_cifar10}
\begin{tabular}{ccccc}
\hline\noalign{\smallskip}
  & \multicolumn{4}{c}{Accuracy}\\
    & Synthetic Images  & Synthetic Images & Synthetic Images & Original Images\\
Model    & $k=250$ & $k=500$ & $k=1000$ & \\
\noalign{\smallskip}
\hline
\noalign{\smallskip}
ResNet20   & 58.00\%  & 61.16\% & 61.79\% & 92.69\% \\
\hline
\end{tabular}
\end{center}
\end{table}
\vspace{-50pt}
\setlength{\tabcolsep}{1.4pt}

\section{Conclusions}
In this work, we present a method for generating synthetic images that can be used for privacy-preserving ML applications. Our method makes use of the BN statistics of the network pre-trained on the original images and optimizes random noise to generate synthetic data that, when fed into the network, have a layer-wise statistical distribution that closely matches the pre-trained model. We evaluate our method on CIFAR10 datasets. Our experiments suggests that model trained from scratch on synthetic images can obtain reasonable accuracy. Moreover, accuracy-privacy trade-off can be achieved by tuning the number of optimization steps used for synthetic data generation, where longer optimization steps increase accuracy sacrificing privacy. Overall, the results and analyses provide evidences in favor of the potentials of our synthetic datasets generation method for privacy-preserving ML applications.          

\section*{Acknowledgements}
This work was supported in part by the Center for Brain Inspired Computing
(C-BRIC), one of the six centers in Joint University Microelectronics Program
(JUMP), in part by the Semiconductor Research Corporation (SRC) Program
sponsored by Defense Advanced Research Projects Agency (DARPA), in part
by the Semiconductor Research Corporation, in part by the National Science
Foundation; in part by the Department of Defense (DoD) Vannevar Bush
Fellowship, in part by the U.S. Army Research Laboratory, and in part by
IARPA.

\bibliographystyle{splncs04}
\bibliography{egbib}

\end{document}